\documentstyle[prc,aps,preprint]{revtex}
\begin{document} \draft
\date{\today}
\title{The coupling constant g$_{\rho\sigma\gamma}$ as derived from QCD sum rules}

\author{A. Gokalp~\thanks{agokalp@metu.edu.tr} and
        O. Yilmaz~\thanks{oyilmaz@metu.edu.tr}}
\address{ {\it Physics Department, Middle East Technical University,
06531 Ankara, Turkey}}
\maketitle

\begin{abstract}
We employ QCD sum rules to calculate the coupling constant
g$_{\rho\sigma\gamma}$ by studying the three point
${\rho\sigma\gamma}$-correlation function. Our results is
consistent with the value of this coupling constant obtained using
vector meson dominance of the electromagnetic current and the
experimental $\rho^0$-photoproduction data.
\end{abstract}

\thispagestyle{empty} ~~~~\\ 
\newpage
\setcounter{page}{1}
\section{Introduction}

The existence of the scalar-isoscalar $\sigma$ meson as a broad
$\pi\pi$ resonance has long been controversial. Recently, an
increasing number of theoretical and experimental analyzes point
toward the existence of this important meson \cite{R1}. Most of
these analyzes find a $\sigma$-pole position near 500-i250 MeV
\cite{R2}. A direct experimental evidence seems to emerge from the
D$^{+}\rightarrow\sigma\pi^{+}\rightarrow 3\pi$ decay channel
observed by the Fermilab E791 collaboration, where $\sigma$ meson
is seen as a clear dominant peak with M$_{\sigma}$=478 MeV and
$\Gamma_{\sigma}$=324 MeV \cite{R3}. Since $\sigma$ meson is a
relevant hadronic degree of freedom, it must be incorporated into
the analysis of hadronic processes.

Although at sufficiently high energies and low momentum transfers
electromagnetic production of vector mesons on nucleon targets has
been explained by Pomeron exchange models, at low energies near
threshold scalar and pseudoscalar meson exchange mechanisms become
important \cite{R4}. In particular, the presently available data
on the photoproduction of $\rho^{0}$ meson on proton targets near
threshold can be described by a simple one-meson exchange model
\cite{R5}. In this picture, $\rho^{0}$ meson photoproduction cross
section on protons is given mainly by $\sigma$ meson exchange. An
important physical input for these studies is the coupling
constant g$_{\rho\sigma\gamma}$. This coupling constant was
estimated by calculating the $\rho\sigma\gamma$-vertex assuming
vector meson dominance of the electromagnetic current and then
performing a fit to the experimental $\rho^0$-photoproduction
data. When this result was derived using an effective Lagrangian
for the $\rho\sigma\gamma$-vertex the value
g$_{\rho\sigma\gamma}$=2.71 is obtained for this coupling constant
\cite{R5}. Since it plays an important role in hadronic processes
involving $\sigma$ meson, it is of interest to study this coupling
constant from a different perspective other than vector meson
dominance.

In this work, we estimate the coupling constant
g$_{\rho\sigma\gamma}$ by employing QCD sum rules which provide an
efficient method to study many hadronic observables, such as decay
constants and form factors, in terms of nonperturbative
contributions proportional to the quark and gluon condensates
\cite{R6,R7,R8}. In section 2, we give a QCD sum rules analysis of
scalar current, derive a sum rule for the overlap amplitude
$\lambda_{\sigma}$ of the scalar current to the $\sigma$ meson
state and estimate this amplitude  since it is not available
experimentally. In section 3, we derive a sum rule for the
coupling constant g$_{\rho\sigma\gamma}$ and by utilizing the
experimental value of $\lambda_{\rho}$, the calculated value of
$\lambda_{\sigma}$ and the known values of condensates we estimate
this coupling constant. Let us here note that the nature of
$\sigma$ meson, whether it is a $\overline{q}q$ state or it is of
another quark, gluon structure is still a matter of debate
\cite{R1,R2,R9}.

\section{QCD sum rules analysis of scalar current}

The QCD sum rules approach  \cite{R6,R7,R8} is a model independent
method to study the properties of hadrons through correlation
functions of appropriately chosen currents. In the SU(2) flavour
limit m$_{u}=m_{d}=m_{q}$, we study the scalar-isoscalar mesons by
considering the scalar current correlation function
\begin{equation}\label{e1}
  \Pi(p^{2})=i\int d^{4}x e^{ip.x}<0|T\{j_{s}(x)j_{s}(0)\}|0>
\end{equation}
where $j_{s}=\frac{1}{2}(\overline{u}u+\overline{d}d)$ is the
interpolating current for $\sigma$ meson. The two-loop expression
for the scalar current correlation function $\Pi(p^2)$ in
perturbative QCD was calculated \cite{R10}, and it is given by the
expression
\begin{equation}\label{e2}
  \Pi_{pert}(p^{2})=\frac{3}{16\pi^2}(-p^2)\ln
  (\frac{-p^2}{\mu^2})\left\{ 1+\frac{\alpha_{s}}{\pi}\left [\frac{17}{3}-\ln
  (\frac{-p^2}{\mu^2})\right ]\right\}~~.
\end{equation}
QCD-vacuum condensate  contributions to the scalar current
correlation function $ \Pi(p^{2})$ were obtained by the operator
product method  \cite{R11}, and they were found in the $m_q$=0
limit as
\begin{equation}\label{e3}
  \Pi(p^{2}=-Q^2)_{cond}=\frac{3}{2Q^2}<m_q\overline{q}q>
  +\frac{1}{16\pi Q^2}<\alpha_s G^2>-
  \frac{88\pi}{27 Q^4}<\alpha_s(\overline{q}q)^2>~~.
\end{equation}
We note that the term $<m_q\overline{q}q>$ is not considered in
the order m$_q$ terms, as this condensate's magnitude, which is
obtained by making use of Gell-Mann-Oakes-Renner relation as
$-f_{\pi}^{2}m_{\pi}^{2}/4$, is independent of quark mass
\cite{R6}.

On the other hand, the correlation function $ \Pi(p^{2})$
satisfies the standard subtracted dispersion relation \cite{R6}
\begin{equation}\label{e4}
  \Pi_{pert}(p^{2})=p^2\int_0^\infty\frac{ds}{s(s-p^2)}\rho(s)+\Pi(0)
\end{equation}
where the spectral density function is given as
$\rho(s)=(1/\pi)Im\Pi(s)$. We parameterize the spectral density as
a single sharp pole $\pi\lambda_\sigma\delta(s-m_\sigma^2)$, where
the overlap amplitude is defined as
$\lambda_\sigma=<0|j_s|\sigma>$, plus a smooth continuum
representing the higher states. The continuum contribution to the
spectral density is estimated in the form
$\rho=\rho_h(s)\theta(s-s_0)$ where s$_0$ denote the continuum
threshold and $\rho_h$ is given by the expression
$\rho_h(s)=(1/\pi)Im \Pi_{OPE}(s)$ with $ \Pi_{OPE}(s)$ obtained
from Eq. (2) and Eq. (3) as
$\Pi_{OPE}(s)=\Pi_{pert}(s)+\Pi_{cond}(s)$. After performing the
Borel transformation we obtain the following QCD sum rule for the
overlap amplitude  $\lambda_\sigma$ of the scalar current
\begin{eqnarray}\label{e5}
  \lambda_{\sigma}e^{-\frac{m_\sigma^2}{M^2}}=\frac{3}{16\pi^2}M^2\left\{
  \left [1-(1+\frac{s_0}{M^2})e^{\frac{-s_0}{M^2}}\right ](1+\frac{\alpha_{s}(M)}{\pi}\frac{17}{3}-
  \frac{\alpha_{s}(M)}{\pi}\int_0^{s_0/M^2} w\ln we^{-w}dw\right\}
 \nonumber \\
  + \frac{3}{2M^2}<m_q\overline{q}q>+\frac{1}{16\pi M^2}<\alpha_s G^2>-
  \frac{88\pi}{27 M^4}<\alpha_s(\overline{q}q)^2>~~.
\end{eqnarray}
For the numerical evaluation of the above sum rule, we use the
values $<m_q\overline{q}q>=(-0.82\pm 0.10)\times 10^{-4}~~GeV^4$,
$<\alpha_s G^2>=(0.038\pm 0.011)~~GeV^4$,
$<\alpha_s(\overline{q}q)^2>=-0.18\times 10^{-3}~~GeV^6$
\cite{R8,R12}, and m$_\sigma$=0.5 GeV. For the continuum threshold
we choose s$_0$=1.1, 1.2, 1.3 GeV$^2$ and we study the M$^2$
dependence between 0.5 GeV$^2$ and 1.4 GeV$^2$. The overlap
amplitude as a function of M$^2$ for different values of s$_{0}$
is shown in Fig. 1 from which by choosing the middle value
$M^2=0.9~~GeV^2$ in its interval of variation we obtain the
overlap amplitude as $\lambda_\sigma=(0.12\pm 0.01)~~GeV^2$ where
we include the uncertainty due to the variation of the continuum
threshold and the Borel parameter $M^2$. Other sources of
uncertainty are the errors attached to the estimated values of
condensates as quoted above. If we take these errors into account
in our analysis as well, we then obtain the value
$\lambda_\sigma=(0.12\pm 0.03)~~GeV^2$ for the overlap amplitude.

\section{QCD sum rules for the $\rho\sigma\gamma$-vertex function}

In order to derive the QCD sum rule for the coupling constant
g$_{\rho\sigma\gamma}$, we consider the three point correlation
function
\begin{equation}\label{e6}
  T_{\mu\nu}(p,p^\prime)=\int d^{4}x d^4y e^{ip^\prime.y}e^{-ip.x}
  <0|T\{j_\mu^\gamma(0)j_{\nu}^\rho(x)j_{s}(y)\}|0>~~.
\end{equation}
The interpolating current for $\rho$ meson \cite{R6} is
$j_{\nu}^{\rho}=\frac{1}{2}(\overline{u}\gamma_{\nu}u-\overline{d}\gamma_{\nu}d)$,
and
$j_{\mu}^{\gamma}=\overline{u}\gamma_{\nu}u+\overline{d}\gamma_{\nu}d$,
$j_{s}=\frac{1}{2}(\overline{u}u-\overline{d}d)$ are the
electromagnetic and scalar currents, respectively.

The theoretical part of the sum rule is obtained by calculating
the perturbative contribution and the power corrections from
operators of different dimensions to the three point correlation
function $T_{\mu\nu}$. For the perturbative contribution we
consider the lowest order bare-loop diagram. Furthermore, we
consider the power corrections from the operators of different
dimensions, $<\overline{q}q>$, $<\sigma\cdot G>$ and
$<(\overline{q}q)^2>$. We do not consider the gluon condensate
contribution proportional to  $<G^2>$ since it is estimated to be
negligible for light quark systems. We perform the calculations of
the power corrections in the fixed point gauge \cite{R13}. We work
in the limit  m$_q$=0, and in this limit perturbative bare-loop
diagram does not make any contribution. Moreover, in this limit
only operators of dimensions d=3 and d=5 make contributions that
are proportional to $<\overline{q}q>$ and  $<\sigma\cdot G>$,
respectively. The relevant Feynman diagrams for the power
corrections are shown in Fig. 2.

In order to obtain the phenomenological part, we consider a double
dispersion relation for the vertex function $T_{\mu\nu}$, and we
saturate this dispersion relation by the lowest lying meson states
in the vector and scalar channels. We obtain this way for the
physical part
\begin{equation}\label{e7}
  T_{\mu\nu}(p,p^\prime)=\frac{<0|j_{\nu}^\rho|\rho>
  <\rho(p)|j_\mu^\gamma|\sigma(p^\prime)><\sigma|j_{s}|0>}
  {(p^2-m^2_\rho)({p^\prime}^2-m^2_\sigma)}+...
\end{equation}
where the contributions from the higher states and the continuum
is denoted by dots. In this expression,
$\lambda_\sigma=<\sigma|j_{s}|0>$ has been determined in Section
2. The overlap amplitude $\lambda_\rho$ of rho meson is defined as
$<0|j_{\nu}^\rho|\rho>=\lambda_\rho u_\nu$ where $u_\nu$ is the
polarization vector of the $\rho$ meson. The matrix element of the
electromagnetic current is given as
\begin{equation}\label{e8}
<\rho(p)|j_\mu^\gamma|\sigma(p^\prime)>=
-i\frac{e}{m_\rho}g_{\rho\sigma\gamma}K(q^2)(p\cdot k ~u_\mu
-u\cdot k~ p_\mu)
\end{equation}
where $q=p-p^\prime$ and $K(q^2)$ is a form factor with K(0)=1.
This expression defines the coupling constant
g$_{\rho\sigma\gamma}$ through the effective Lagrangian
\begin{equation}\label{e9}
{\cal L}=\frac{e}{m_{\rho}}g_{\rho\sigma\gamma}\partial^\alpha
\rho^\beta(\partial_\alpha A_\beta-\partial_\beta A_\alpha )\sigma
\end{equation}
describing the $\rho\sigma\gamma$-vertex \cite{R5}.

After performing double Borel transform with respect to the
variables $Q^2=-p^2$ and ${Q^\prime}^2=-{p^\prime}^2$, and by
considering the gauge-invariant structure ($p_\mu q_\nu -p\cdot q
g_{\mu\nu}$) we obtain the sum rule for the coupling constant
g$_{\rho\sigma\gamma}$
\begin{eqnarray}\label{e10}
g_{\rho\sigma\gamma}=\frac{3m_\rho}{\lambda_\rho\lambda_\sigma}
  e^{\frac{m_\rho^2}{M^2}}e^{\frac{m_\sigma^2}{{M^\prime}^2}}<\overline{u}u>
    \left (-\frac{3}{4}+\frac{5}{32}m_0^2\frac{1}{M^2}
    -\frac{3}{32}m_0^2\frac{1}{{M^\prime}^2}\right )
\end{eqnarray}
where we use the relation $<\sigma\cdot G>=m_0^2<\overline{q}q>$.
For the numerical evaluation of the sum rule we use the values
$m_0^2=(0.8\pm 0.02)~~GeV^2$, $<\overline{u}u>=(-0.014\pm
0.002)~~GeV^3$ \cite{R8,R14}, and $m_\rho=0.77~~GeV$,
$m_\sigma=0.5~~GeV$. For the overlap amplitude $\lambda_\sigma$ we
use the value $\lambda_\sigma=(0.12\pm 0.03)~~GeV^2$ that we have
estimated in Section 2 and for $\lambda_\rho$ we use its
experimental value $\lambda_\rho=(0.107\pm 0.003)~~GeV^2$ as
obtained from the measured leptonic width $\Gamma
(\rho^0\rightarrow e^+ e^-)$ of $\rho$ meson \cite{R15}. In order
to analyze the dependence of g$_{\rho\sigma\gamma}$ on Borel
parameters $M^2$ and ${M^\prime}^2$, we study the independent
variations of $M^2$ and ${M^\prime}^2$ in the interval
$0.6~~GeV^2\leq M^2,{M^\prime}^2\leq 1.4~~GeV^2$ as these limits
determine the allowed interval for the vector channel \cite{R16}.
The variation of the coupling constant g$_{\rho\sigma\gamma}$ as a
function of Borel parameters $M^2$ for different values of
${M^\prime}^2$ is shown in Fig. 3, examination of which indicates
that it is quite stable with these reasonable variations of $M^2$
and ${M^\prime}^2$. We choose the middle value $M^2=1~~GeV^2$ for
the Borel parameter in its interval of variation and obtain the
coupling constant g$_{\rho\sigma\gamma}$ as
g$_{\rho\sigma\gamma}=3.2\pm 0.2$ where only the error arising
from the numerical analysis of the sum rule is considered. The
other sources contributing to the uncertainty in the coupling
constant besides those due to variations of $M^2$ and
${M^\prime}^2$ are the uncertainties in the estimated values of
the vacuum condensates. If we take these uncertainties into
account, we then obtain the coupling constant
g$_{\rho\sigma\gamma}$ as g$_{\rho\sigma\gamma}=3.2\pm 0.6$.

Our estimate of the coupling constant g$_{\rho\sigma\gamma}$ is in
agreement with its value deduced from the analysis using the
vector meson dominance of the electromagnetic current for the
$\rho\sigma\gamma$-vertex. On the other hand, an independent
anaysis \cite{R17} utilizing the experimental value of the decay
rate $\Gamma (\rho^0\rightarrow \pi^+\pi^-\gamma)$ and using the
values for the $\sigma$ meson parameters $m_\sigma$=478 MeV and
$\Gamma_\sigma$=324 MeV that are obtained from the Fermilab E791
experiment gives the value g$_{\rho\sigma\gamma}=5.92\pm 1.34$ for
this coupling constant. However, we like to note that in the
present work we consider $\sigma$ meson in the narrow resonance
limit and do not take the finite width of $\sigma$ meson into
account.

\newpage
\begin{center}
{\bf ACKNOWLEDGMENTS}
\end{center}

We thank  T. M. Aliev for helpful discussions during the course of
our work.


\newpage

{\bf Figure Captions:}

\begin{description}

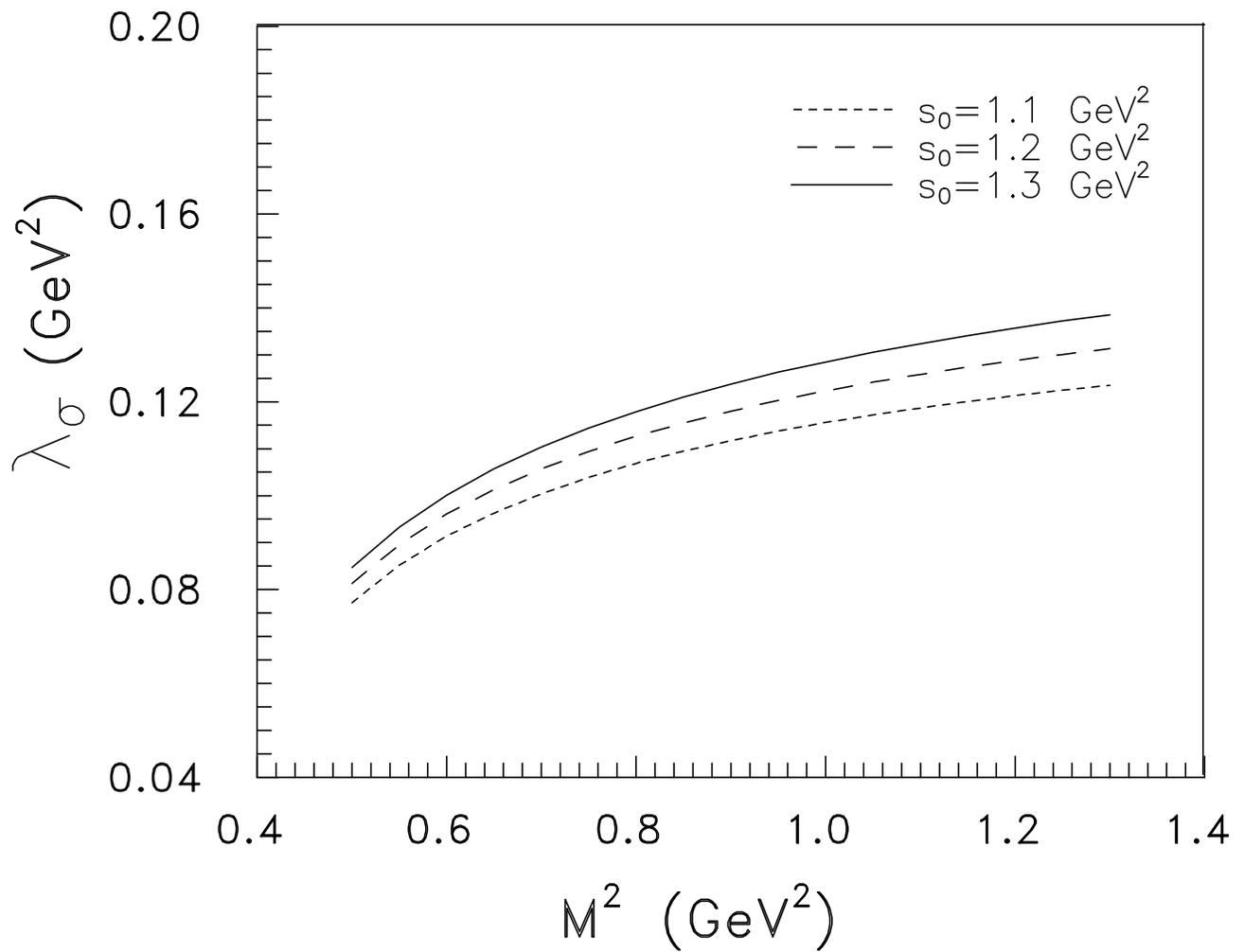
\item[{\bf Figure 1}:] The overlap amplitude $\lambda_\sigma$ as a
function of Borel parameter $M^2$.

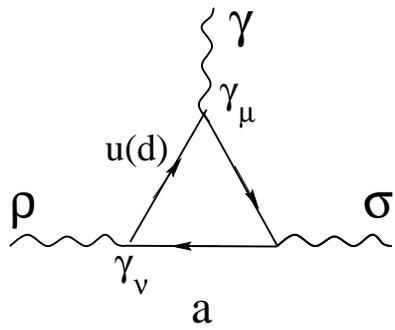
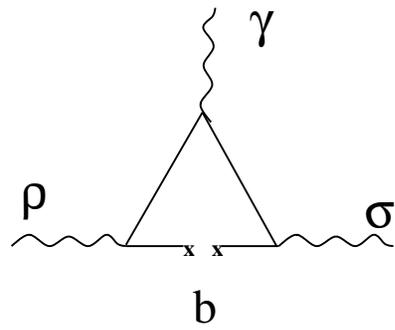
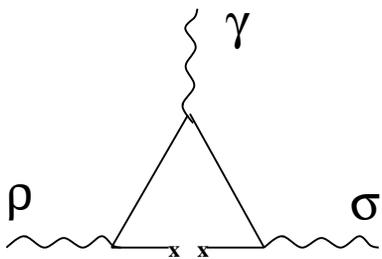
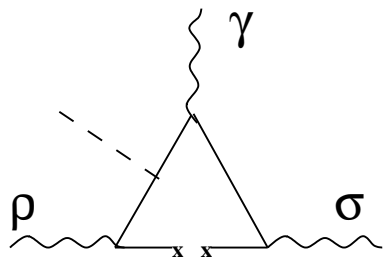
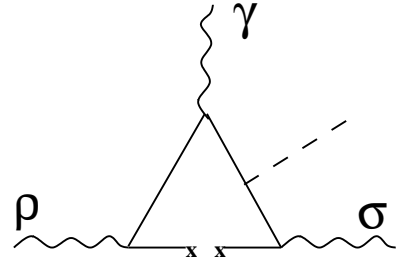
\item[{\bf Figure 2}:] Feynman Diagrams for the
$\rho\sigma\gamma$-vertex: a- bare loop diagram, b- d=3 operator
corrections, c- d=5 operator corrections. The dotted lines denote
gluons.

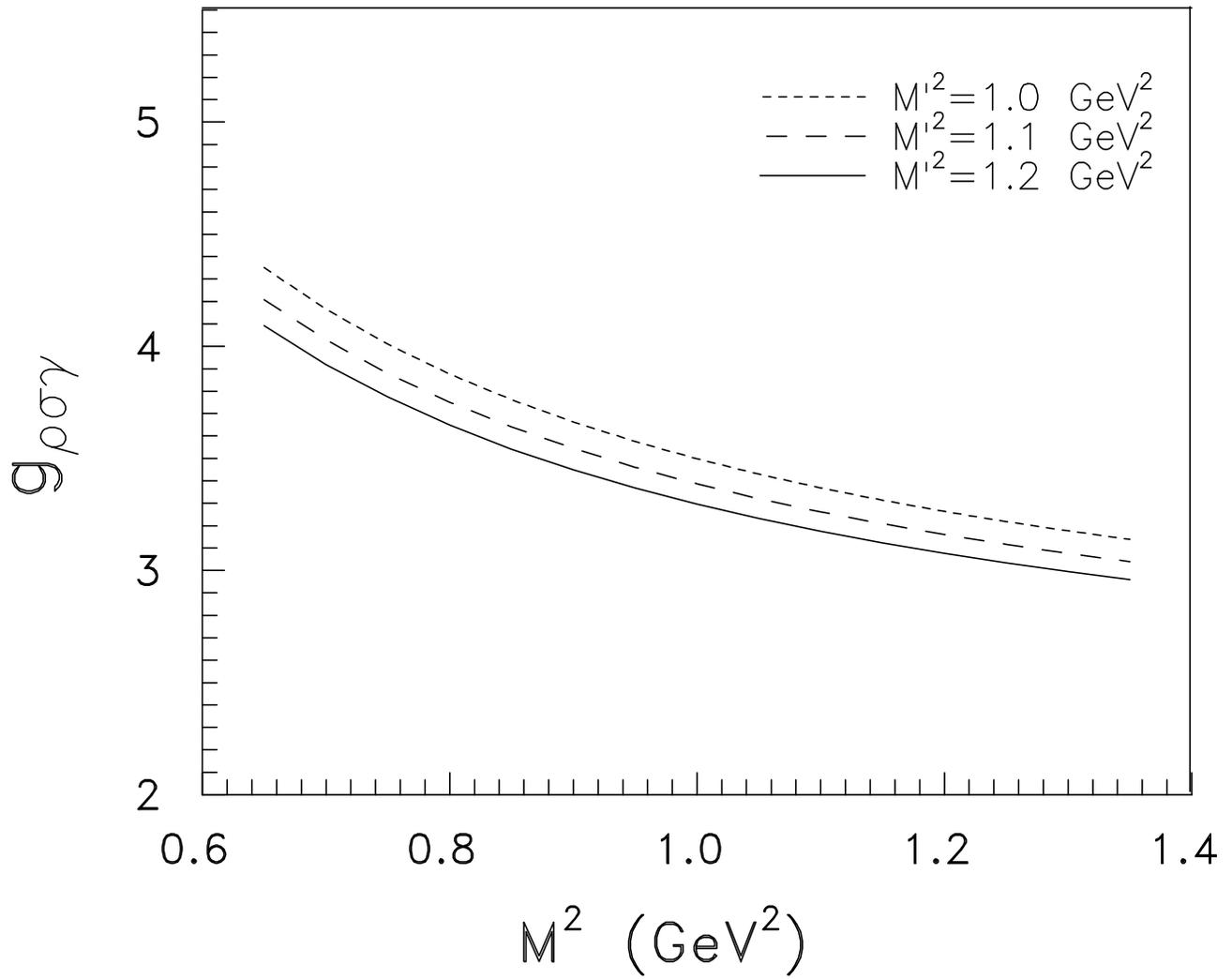
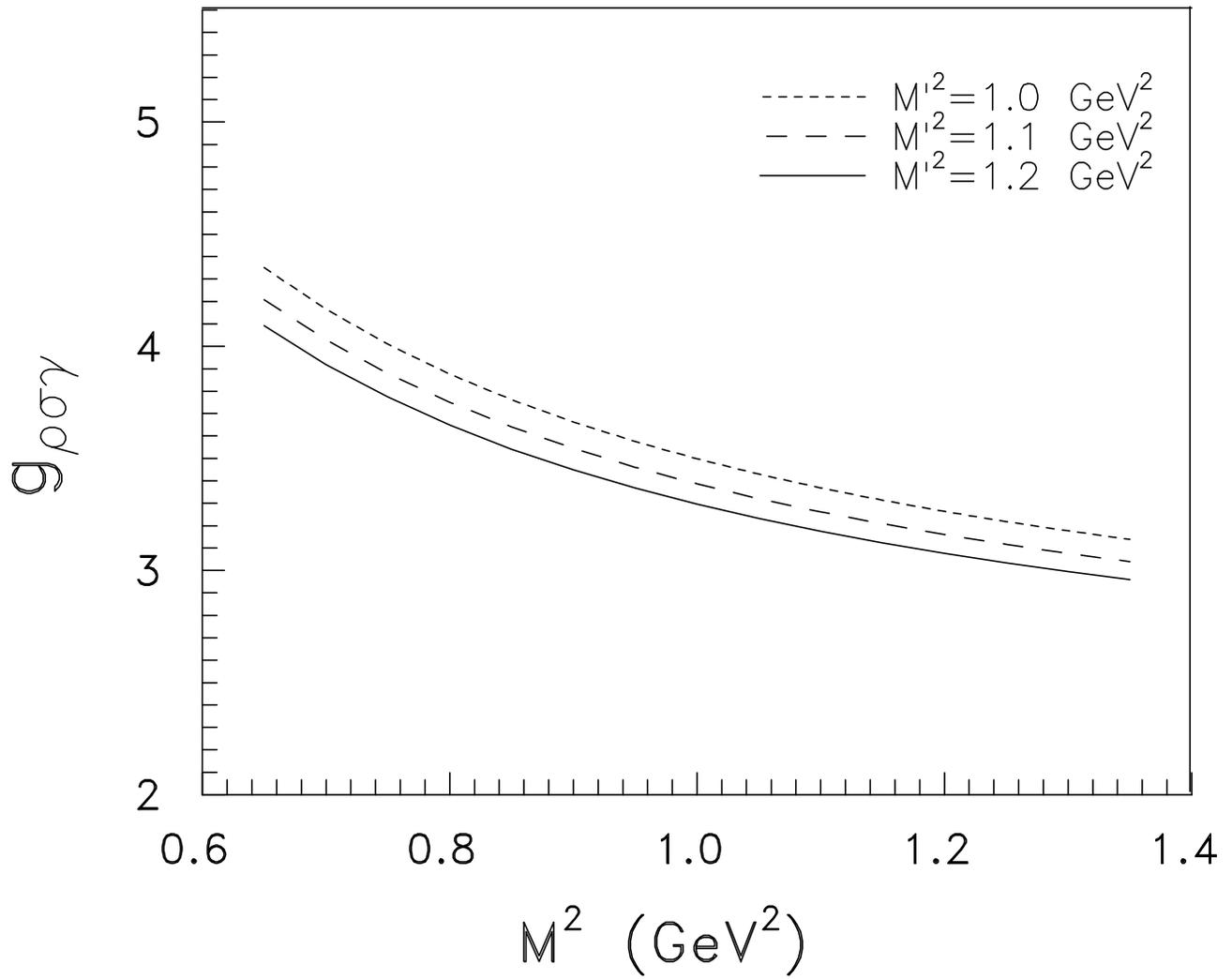
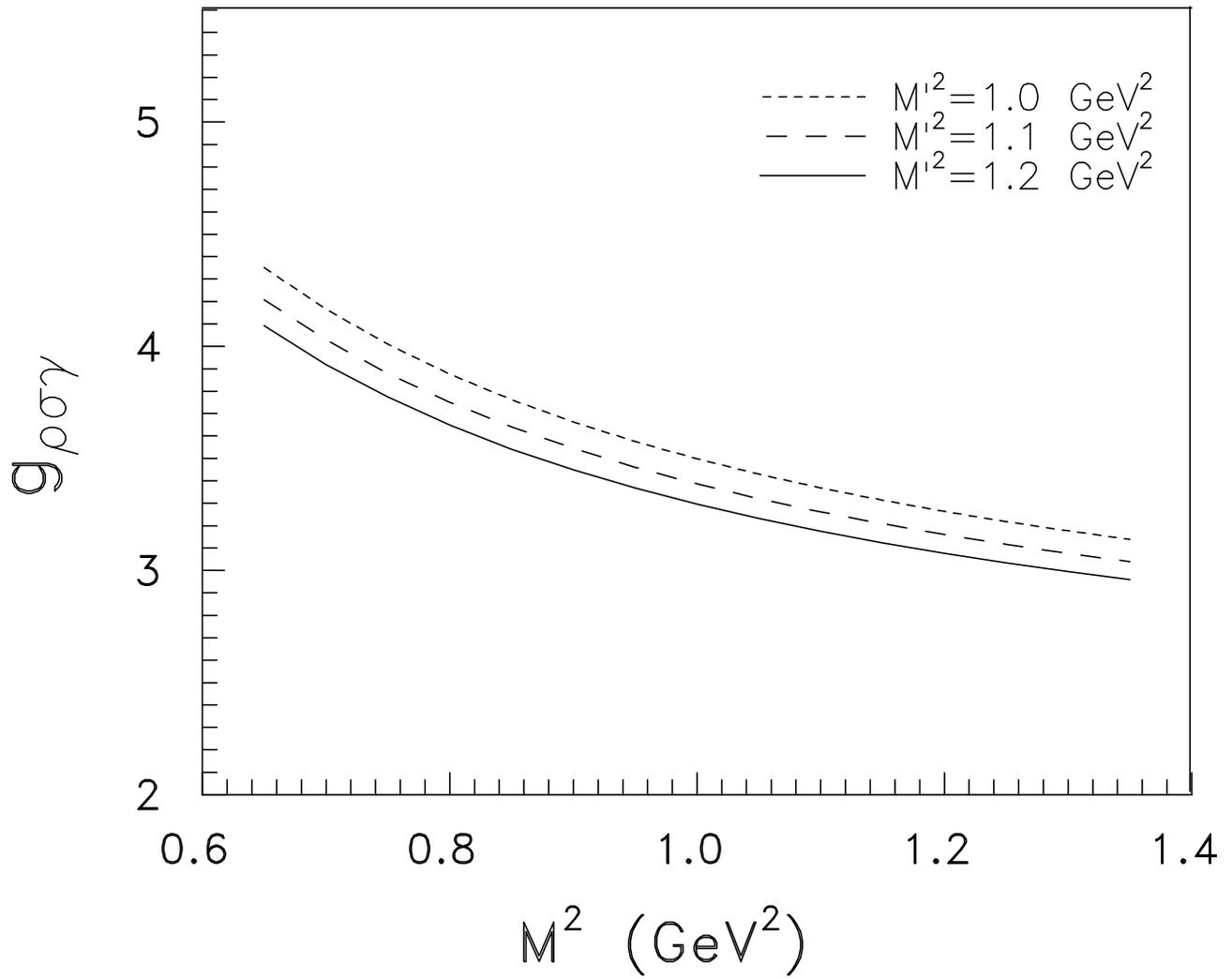
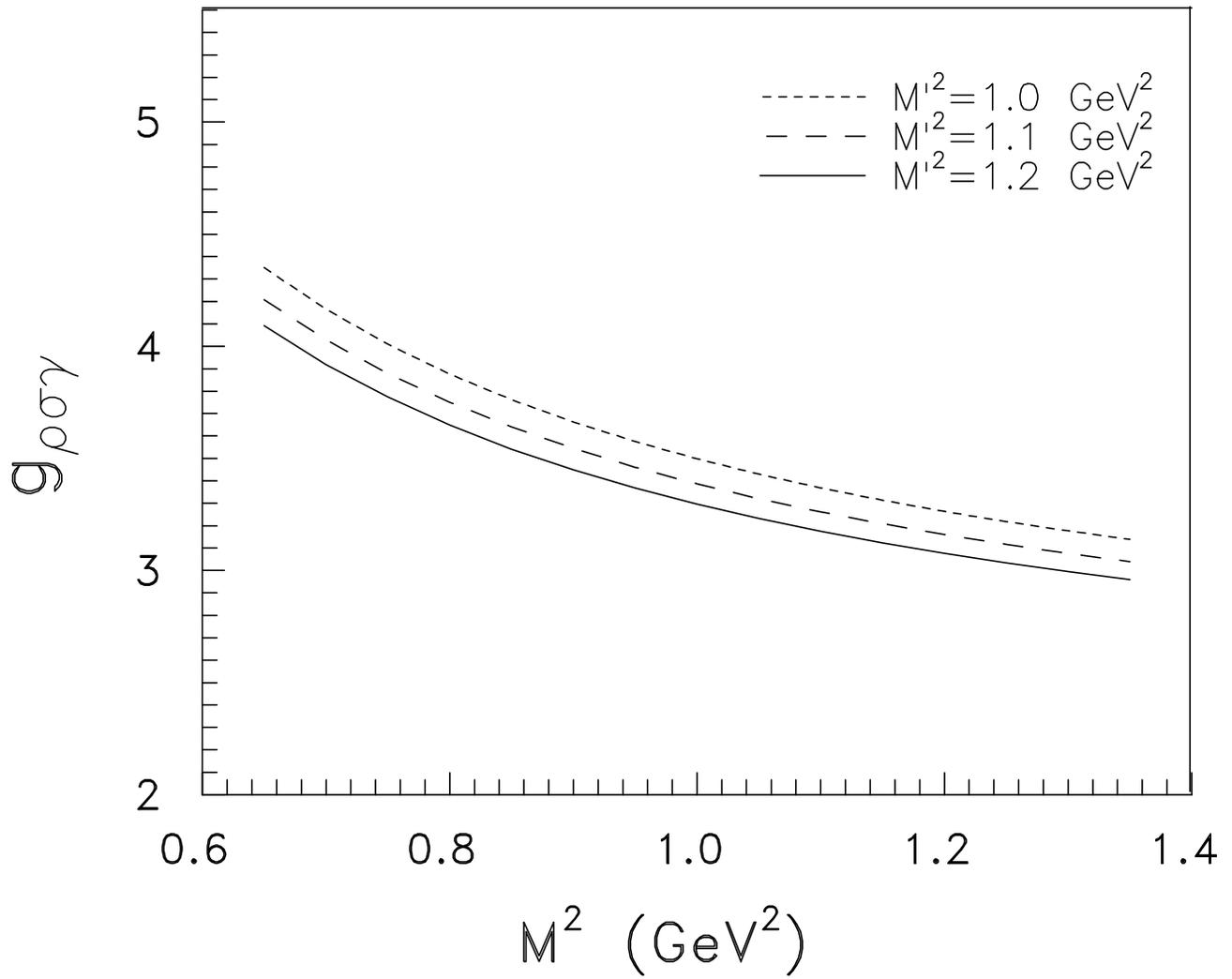
\item[{\bf Figure 3}:] the coupling constant $g_{\rho\sigma\gamma}$
as a function of the Borel parameter $M^2$ for different values of
${M^\prime}^2$.

\end{description}

\end{document}